\documentclass[12pt,twocolumn,twoside]{article}

\usepackage{authblk}
\usepackage[left=15mm,top=0.5in,right=15mm,bottom=1in]{geometry}
\usepackage{abstract}
\usepackage{graphicx}
\usepackage{wrapfig}
\usepackage{lscape}
\usepackage{rotating}
\usepackage{epstopdf}
\usepackage{hyperref}
\usepackage{amsmath}
\usepackage{amssymb}
\usepackage[noadjust]{cite}

\title{Monotonous manipulation of atomic density in an isovolumetric focused-beam trap}
\author[1]{\large Nuttanan Tanasanchai}
\author[1]{\large Kritsana Srakeaw}
\author[1]{\large Jindaratsamee Phrompao}
\author[1,2,*]{\large Waranont Anukool}
\affil[1]{\small \textit{Department of Physics and Materials Science, Faculty of Science, Chiang Mai University, 239 Huay Keaw Road, Muang, Chiang Mai, 50200, Thailand}}
\affil[2]{\small \textit{Thailand Center of Excellence in Physics, Commission on Higher Education, 328 Si Ayutthaya Road, Bangkok 10400, Thailand}}
\affil[*]{\small \textit{Corresponding author: waranont.a@cmu.ac.th}}

\date{}
\title{Monotonous Manipulation of Atomic Density in an Isovolumetric Focused-beam Trap}

\begin{document}
\twocolumn[
%\begin{@twocolumnfalse}
	\maketitle
	\renewcommand{\abstractname}{}    % clear the title
	\vspace*{-2cm}
	\begin{abstract}
		We present an asymmetric focused-beam trap of rubidium-85 atoms where the cloud volume only varied with the trap laser intensity. By repumping on F=2 to F'=2 D2 transition, the Stokes Raman scattering has strongly affected the population ratio in the two hyperfine ground states. With the volume and trap number simultaneously and separately controlled, we have shown that the cloud density can be monotonously manipulated from zero to the maximum value limited by the cooling power. By solely varying the intensity of repump laser, the scheme for precision loading of exact 1-6 atoms in the FORT, which is beyond the practical loading limit of the blue-detuned light-assisted collision, is described.
	\end{abstract}
%\end{@twocolumnfalse}
]

A few years after the first Magneto-Optical Trap (MOT) had been demonstrated in 1987 \cite{1}, several succeeding trap designs have been proposed and verified with particular properties departing from the six-beamed MOT. They may be classified by mechanical protocols \cite{2, 3}, and magneto-optical conformation \cite{4,5,6,7,8}. Though the MOT is still most widely employed to prepare cold dense atomic cloud from room-temperature vapor, today's diverse applications demand traps with rather specific characteristics. 

Aiming at neutral-atom precision loading of the far-off-resonance optical dipole trap (FORT) for cavity QED experiments, the two-beamed magneto-optical trap (TBT)  \cite{9}, where transverse momentum exchange arises as a result of focused cooling beams, has been modified and investigated. Our asymmetric focused-beam trap (FBT) of rubidium-85 atoms comprises two confocal cooling beams in the z-axis and two normal pairs of counter-propagating repump beams on the transverse plane. Besides F=2 to F'=2 transition for repumping, the opto-magnetic configuration is similar to the standard MOT. We have characterized the FBT to optimize independent parameters, i.e. detuning and intensities of trap beams, magnetic field gradient, and background pressure, for the highest trap number. After that the relations among the cloud density, number of atoms, and the total intensity of repump beams were studied within two regimes. First is the multi-scattering regime, where the fluorescence photons get reabsorbed and radiative pressure limits the density. Second is the two-component regime, where the photon re-absorption spills out atoms from the harmonic trap and hence lessens the cloud density.

The quantitative analyses of the FBT manifest a uniform cloud volume over the whole experimental range of atomic density and repump laser intensity. The cloud volume was controlled by varying the trap laser intensity. Within both regimes, we have demonstrated monotonous manipulation of the cloud density through varying only the intensity of the repump beams. Rough estimation of the repumper intensities for loading 1-6 atoms in the FORT is also given. 

Particularly for the focused-beam configuration where the trap frequency is detuned by a few MHz from an atomic cyclic transition, the associated pair of light dressed states is strongly coupled and the AC Stark Effect \cite{10} responsible for spatially dependent energy shifts is largely suppressed. Without the effect of light shift considered, the optical angular momentum induced through the modified wavefront would only throw atoms around. For an atom moving toward the trap center along the z-axis, the alternate optical resonances of the two velocity components would yield a wavy classical trajectory during the Doppler cooling cycles. The radial momentum transfers are averaged out over many photon absorptions and emissions. If the atom enters the trap region on the transverse plane, the Doppler cooling reclaims its effectiveness only on the outbound course because of the instantaneous vectorial mismatch between the atomic velocity and the spatially dependent wave vector of the focused lasers. All in all, the FBT could still provide three-dimensional cooling and confinement owing to the transverse momentum exchanges through atom-light scattering process.

In this work, all self-made external cavity diode lasers employed cateye configuration \cite{11}. For the cooling beams, the linearly-polarized laser (diode: DL-780-90) was divided with a polarization beam splitter (PBS) into 68 mW and 2 mW of power. The first one was locked at 5 MHz below the F=3 to F'=(3,4) crossover transition by using the feedback signal from the second weaker beam that was sent to the single-beamed Doppler-free saturated absorption spectroscopy (DSAS) \cite{12}. The cooling beam was injected into a double-pass 80 MHz acousto-optic modulator (AOM). It was then amplified at a tapered chip (EYP-TPA-0780-01000-3006-CMT03-0000) before doubly passing through a 110 MHz AOM. This section permits both varying the intensity and shifting the frequency within a range of 20 - 100 MHz from the atomic transition as mentioned. Single-mode optical fibers were used as laser waveguides that separated the beam preparation from the experimental space around the rectangle rubidium cell ($1.00~{\rm cm} \times 2.00~{\rm cm} \times 9.75 ~{\rm cm}$ of inside dimensions). Right after the outward-bound fiber coupler, the cooling laser was expanded, equally split and polarized into two $\sigma^+$ and $\sigma^-$ Gaussian beams. Identically on both sides, a 32 mm-focal-length lens and a 0.76 NA lens (ACL50832U-B) were combined to tightly focus each cooling beam before counter-propagating with the other to form a confocal arrangement along the z-axis at Fourier plane separation of 1.00 cm.

During the experiment, the repump laser (diode: L785P100) was kept locked using similar DSAS configuration at the frequency 2.7 MHz above the F=2 to F'=(1,2) crossover transition. A pair of half-wave plate and PBS helped to attenuate the power of the repump beam before getting expanded from 1.62 mm to 6.82 mm of diameter. These beams provide the overlap volume well covering the whole cooling region. The repump laser was split and circularly polarized in the same manner as the cooling one. Two normal pairs of counter-propagating repump beams lying on the xy-plane intersect the trap beams at the center of the borosilicate glass cell coinciding with the zero magnetic field from the anti-Helmholtz coils.

The experiment began with an ergonomic optimization of the FBT to find a set of independent parameters close to the maximal efficiency pertaining to the highest trap number $(N)$. The optimal magnetic field gradient $(\nabla {\bf B})$, background pressure $(P)$, detuning $(\delta _c)$ and intensities $(I_c)$ of cooling beams were determined and concluded in Table \ref{tab:parameters}. According to the broad plateau of the cloud density $(\rho)$ vs. $\nabla {\bf B}$  observed beyond $15.0~{\rm G\cdot cm^{-1}}$, we only executed experiments at $16.48~{\rm G\cdot cm^{-1}}$. In order to verify that  $I_c=4.95~{\rm mW\cdot cm^{-2}}$ is well within the multi-scattering regime, a series of experiments for $\rho$ vs. $I_c$ has been conducted at four fixed values of repump intensities, i.e. $I_r= 0.33,0.62,2.48$ and $17.48~{\rm mW\cdot cm^{-2}}$. With all parameters defined, the physical connections among  $\rho, N,$ and $I_r$ can be investigated by varying the repumper intensity from  $0.21-17.85~{\rm mW\cdot cm^{-2}}$. Three $\rho$ vs. $I_r$ experiments performed at $I_c>4.95~{\rm mW\cdot cm}^{-2}$ are to confirm the consistency of parametrical setup.

 \begin{table}[htbp]
 	\centering
 	\caption{Independent parameters optimized for this work.}
 	\begin{tabular}{lcc}
 		\hline
 		Control Parameters & $N$ vs. $I_c$ & $N$ vs. $I_r$ \\
 		\hline
 		$I_c ~~(\rm mW\cdot cm^{-2})$ & $1.15 - 10.36 $ & $4.95 - 9.92 $ \\
 		$\delta_c~~(\rm MHz)$ & $-12$ & $-12$ \\
 		$I_r~~(\rm mW\cdot cm^{-2})$ & $0.33-17.48$ & $0.21-17.85$ \\ 
 		$\delta_r~~(\rm MHz)$ & $-12$ & $-12$ \\ 
 		$\nabla {\bf B}~~(\rm G\cdot cm^{-1})$&  $16.48$ & $16.48$ \\ 
 		$P ~~(\rm mbar)$  & $  6.04 \times 10^{-9}$ & $5.83 \times 10^{-9}$ \\ 
 		\hline
 	\end{tabular}
 	\label{tab:parameters}
 \end{table}

 To get a large damping rate for cooling in the z-axis, we have detuned the trap laser by -12 MHz from F=3 to F'=4 transition which is approximately two times of the decay rate. By slowly changing the external cavity length, the frequency scanning has identified two comparable maxima of the florescence signal at the repumper detuning ($\delta_r$) of -75.4 MHz (F=2 to F'=3) and -12 MHz (F=2 to F'=3); the transition generally engages in the standard MOT. However, -75.4 MHz (F=2 to F'=3) detuning which is equivalent to $\delta_{r}$=-12 MHz (F=2 to F'=2), does not lead to a shared excited state, therefore, the stimulated Raman transition was suppressed. The spontaneous Raman photons from the repump beams were used as the control over the trap number. Taking this into account, we applied -12 MHz (F=2 to F'=2) of detuning to the repump beams, and overruled the information of the bright state population by simultaneously probing the cold atoms in both ground states, i.e. F=2 and F=3, using repump and trap lasers respectively.
 
 The number of rubidium-85 atoms were estimated from pictures of the cold cloud taken while in steady state by using fluorescence imaging technique (EMCCD camera: Andor iXon 897) and Gaussian fitting routine. Since the cloud geometry was best described by an oblate spheroid, the volume of the ensemble was calculated using the radii along two semi axes that correspond to $e^{-1}$ peak signals. The temperature was estimated employing the free expansion method. The studies of the effect of trap beam and repumper intensities were carried out with increasing and decreasing fields respectively while maintaining the trap geometry, beam sizes, and polarizations. Few data points had been repeated to check for parametric consistency and were inclusive in the analyses. The quantities measured were averaged over ten identical experiments successively repeated. All coherent optoelectronic and magnetic operations were controlled by a self-made timing sequence generator with $<100 ~ \mu s$ of resolution. Unless otherwise stated, physical values in the following discussion are related to $I_c = 4.95~{\rm mW\cdot cm}^{-2}$ and $I_r =17.48~{\rm mW\cdot cm}^{-2}$ that gave the maximum density.

At the optimal control parameters, the harmonic trap depths of the FBT were roughly estimated to $11.04~K$ and $11.73~K$ in the z-axis and on the transverse plane respectively. The gaseous spheroids were flattened along the z-axis by approximately 30\% of the radius from a perfect sphere and the largest volume has semi-axis lengths of 237 $\mu m$ and 310 $\mu m$. While varying the cooling power, the loading times placed between 387~ms to 634~ms were almost independent of the repumper intensity. The lowest temperatures measured of the cloud were $41.7~\mu K$ on the shorter semi-axis and $270.9~\mu K$ on the radial direction. The Doppler temperature of $145.6~\mu K$ for Rb-85 atoms suggests that the polarization gradient cooling was not as effective as in the standard MOT along the symmetry axis and has not been activated on the plane containing minor axes probably due to the heating effect of the Stokes Raman scattering.

\begin{landscape}
	\begin{figure}
	\centering
	\includegraphics[width=1\linewidth]{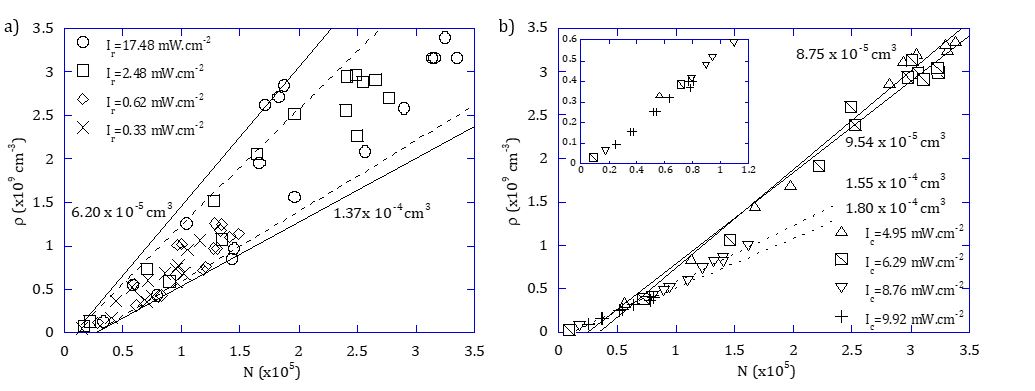}
	\caption[width=\textwidth]{shows the cold cloud density ($\rho$) as a function of the trap number ($N$) when varying a) the trap beam intensity ($I_c$) and b) the repumper intensity ($I_r$). Each fit line in Figure \ref{fig:cold}b gives a constant volume at fixed $I_c$.}
	\label{fig:cold}
\end{figure}
\end{landscape}

The maximum density observed was $3.35\times10^{9}~{\rm cm}^{-3}$ at optimal conditions, i.e. $I_c=4.95~ {\rm mW\cdot cm^{-2}}$, and $I_r=17.48~{\rm mW\cdot cm^{-2}}$, and almost constant within 20 $\mu m$ of radius from the trap center in all directions. At a fixed $I_r$, the trap beam geometry confined atoms within $6.20-13.7\times10^{-5}~{\rm cm}^{3}$ of volume bounded by two solid lines in Figure \ref{fig:cold}a. The behavior of $\rho$ vs. $I_c$ shows non-monotonic relation within the multi-scattering regime which is of interest in this work. We varied $I_r$ to further investigate the FBT in the multi-scattering and two-component regimes with different values of $I_c$ (Figure \ref{fig:cold}b). While the standard MOT demonstrates constant density up to $N=10^{10}$ \cite{13}, the tight confinement of the FBT seems to impose strong restriction on both density and the trap number to undergo changes with $I_c$ at the same rate as illustrated in Figure \ref{fig:cold}b. The linear relationship of $\rho$ vs. $N$ implies constant volume throughout the experiment where four measures of $8.75\times10^{-5}, 9.54\times 10^{-5}, 1.55\times 10^{-4}$ and $1.80\times 10^{-4}~{\rm cm}^{3}$  were observed at  $I_c=$ 4.95, 6.29, 8.64, and 9.97 ${\rm mW\cdot cm}^{-2}$ respectively. This feature also implies that the cloud volume only changed with $I_c$ and not $I_r$. Counting atoms in both hyperfine ground states using the scattering rate of F=3 to F'=4 transition yielded underestimated $N$. However at a constant volume, the linear relation of $\rho$ vs. $N$ remains on the same straight line.

\begin{landscape}
	\begin{figure}
		\centering
		\includegraphics[width=1\linewidth]{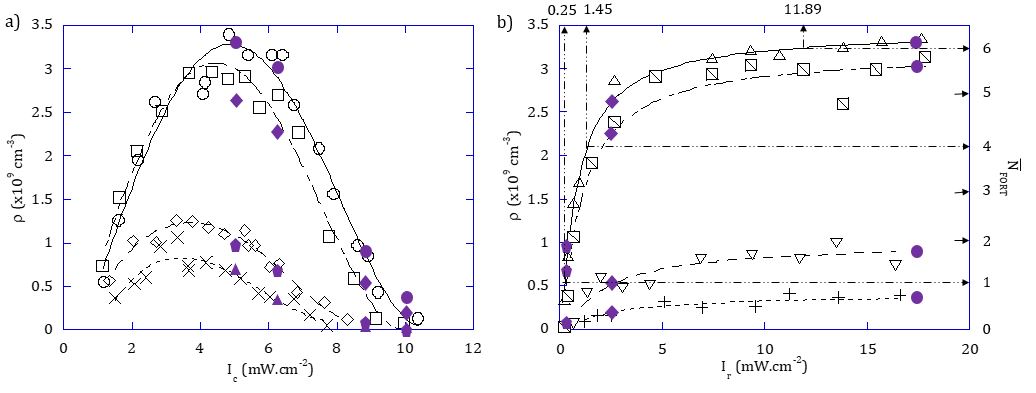}
		\caption[width=\textwidth]{illustrates the dependences of the cloud density ($\rho$) on a) the intensities of the trap laser ($I_c$) and b) the repump laser ($I_r$) using the same symbols as described in Figure\ref{fig:cold}. The same filled symbols indicate the cloud densities at constant $I_r$ while a set of four different filled symbols manifests a constant cloud volume within the same trap regime. All filled symbols were marked on the fit curves in Figure \ref{fig:den-I}b then linked back to Figure \ref{fig:den-I}a for consistency checking. The polynomial fits are for guidance. The rightmost axis shows average number of trapped atoms ($\overline{N}_{FORT}$) expected in the FORT with beam waist of $1.2~\mu m$ at the center of the FBT. Three values of repumper intensity shown topmost are taken from Table \ref{tab:FORT}.}
		\label{fig:den-I}
	\end{figure}
\end{landscape}

The atomic density of the cold cloud strongly depends on the loading rate and the loss rate. The saturated densities in Figure \ref{fig:den-I}a are given by the balance between the confining force and the multiple scattering force. The first one is not harmonic since it is the combination of the damping force on the inbound and restoring force on the outbound of propagating atoms. The second is the competition between the repulsion due to reabsorption of fluorescence photon between atoms and the compression intensified via the shadow effect. The FBT has demonstrated highest density within a narrow window, $3.0<I_c< 6.0~{\rm mW\cdot cm}^{-2}$ (Figure \ref{fig:den-I}a), on the rising leg of $I_c$ before the damped harmonic well spilled out excessive atoms resulting in a sharp drop of  $\rho$. However, the trap beam intensity at $I_c=4.95~{\rm mW\cdot cm}^{-2}$  well maintained the clouds in the multi-scattering regime for the whole range of $I_r$ imposed. The local maxima incline toward increasing $I_c$ from $\rho = 0.83\times 10^{9}~{\rm cm}^{-3}$ to $3.29\times 10^{9}~{\rm cm}^{-3}$ in accordance with an average volume of $9.60\times 10^{-5}~{\rm cm}^{3}$ narrowed down to the middle zone of Figure \ref{fig:cold}a. In the present case, the repump-to-trap intensity ratio has been increased to $>$11 times of the magnitude generally utilized in the standard MOT. The F=2 ground state would be fast depleted while the Stokes Raman transition heated up atoms. As a result, no further compression on the xy-plane was observed which is consistent with the directional measured temperatures. In contrast to the $\rho-I_c$ relation, the cloud density and hence $N$ went up monotonically with $I_r$. It never decreased but smoothly approached the asymptote line corresponding to the density of $3.45\times 10^9~{\rm cm}^{-3}$ at the optimal cooling intensity.

While most physical properties of the FBT resemble typical MOTs, the effect of repump beams on the number of trapped atoms is radically different from the Wieman-Pritchard model used to describe an additional repulsive force arising from multiple reabsorptions of Stokes Raman photons in the optical thickness regime \cite{13}. Consider the standard MOT, the cooling process would get saturated approximately a thousand times slower than the repumper due to small leakage out of the F=3 to F'=4 cooling cycle. This implies that weak repump beams are all the Doppler cooling needs and changing the intensity of the repumper should not affect the density of the cold cloud. On the contrary to the FBT geometry, where cooling along all three principal axes are off balance, the repumper plays a role of diligent trap enhancer that could also completely suppress the trap number even at the most efficient cooling intensity. In this work where the maximum trap number was just $3.35\times 10^5$, by further detuning the repump frequency at -75.4 MHz from the F'=3 hyperfine state, the excited states of the two lasers are well separated and the $I_r-$ dependence of the cloud density got amplified. Evidently, further compression of the cold cloud with growing $I_r$ cannot be understood by integrating the repumper heating of $\hbar(\Delta\omega_{23}-2\pi\times 12$ MHz) per Stokes photon scattering as an additional repulsion force in the rate equation. Though including all parameters that influent the capture velocity is too complex, strong $I_r-$ dependence of the cloud density are expected to behave like $\rho = \rho _{\infty}(\frac{I_r}{I_N+I_r})^2$ directly obtained from a simple rate equation. Here, $\rho _{\infty}$ represents the maximum density and $I_N$ denotes reference intensity relating to the proportion of the hyperfine ground-state populations inside the FBT. We found that the repulsive force due to Stokes photon heating does not overrule the effect of hyperfine ground state conversion to F=3, as a result the density increased with $I_r$. We deduced $I_N = 0.38, 0.49, 1.01,$ and $1.12~{\rm mW\cdot cm}^{-2}$ in descending order of $I_c$ in Figure \ref{fig:den-I}b.

\begin{table}[htbp]
	\centering
	\caption{The intensities of repump laser required for loading 1-6 atoms in the FORT with beam waist of 1.2 $\mu m$.}
	\begin{tabular}{lccc}
		\hline
		\scriptsize
		 $\overline{N}_{\rm FORT} $& \scriptsize$\overline{N}_{FBT} (\times 10^3)$ &\scriptsize $\rho_{\rm FBT} (\times 10^9 {\rm cm}^{-3})$ &\scriptsize $I_r~({\rm mW\cdot cm}^{-2}) $  \\
		\hline
		1 & 48 & 0.54 & 0.25\\
		2 & 97 & $ 1.08 $ & 0.49\\
		3 & 146 &1.62 & 0.82 \\ 
		4 & 194 & 2.16 & 1.45 \\ 
		5 &  243 & 2.70 & 2.93\\ 
		6 & 292 & 3.24 & 11.89\\ 
		7 & 341 & 3.78 & -\\
		\hline
	\end{tabular}
	\label{tab:FORT}
\end{table}

To serve the purpose intended, loading of the FORT from the FBT is straight forward when both traps share the same pair of high NA lenses. Since the minimum waist of the FORT is confined to the diffraction limit of the wavelength used, only tiny overlap volume of the two traps at the cloud center defines the number of atoms that takes part. By using the FBT region with nearly constant density and the FORT with beam waist of 1.2 $\mu m$, the overlap volume has the trap geometry prescribed by 40.0 $\mu m$ and 2.4 $\mu m$ along the axial and radial axes respectively. Precise 1-6 atoms may be transferred into the FORT by varying $I_r$ from $0.25 - 11.89 ~{\rm mW\cdot cm}^{-2}$, where loading 7 atoms in the FORT would require $I_r > 20~{\rm mW\cdot cm}^{-2}$ which is out of the scope of this work. Selectively trapping more than two atoms in the FORT is beyond the practical limit of the blue-detuned light-assisted loading \cite{14,15}. To use the scheme illustrated in Figure \ref{fig:den-I}b, the values of $I_r$ must be evaluated for distinct trap conformation.

With all control parameters held fixed at ergonomically optimized values, our FBT demonstrated  $I_c-$dependent uniform volume of the cold cloud that was not influenced by $\rho$ and $I_r$ up to $3.35\times 10^9 {\rm cm}^{-3}$ and 17.85 ${\rm mW\cdot cm}^{-2}$ respectively. Without varying the intensity of trap beams, the spontaneous Raman photons from an intensified repump laser permits monotonous manipulation of the cloud density from zero to the maximum density limited by the cooling power. Our tight trap arrangement provides an alternative precision loading of the FORT for cavity QED experiments. The plausibility for independent loading of individual FORTs in micro-trap arrays would extend the experimental range of quantum simulation on complex many-body systems that demand selectable degrees of freedom at prescribed lattice defects.

%---------------

The project is funded by Thailand Center of Excellence in Physics, Commission on Higher Education, 328 Si Ayutthaya Road, Bangkok 10400, Thailand.

The authors would like to show gratitude to Prof. Emeritus Dr. Thiraphat Vilaithong and Prof. Dr. Sitthichai Pokai-udom for their continuous support and patience. Deep appreciation is given to all who conducted this experiment. Special acknowledgement goes to National Astronomical Research Institute of Thailand, Mahanakorn University of Technology and Synchrotron Light Research Institute.

\renewcommand\refname{}
\vspace*{-1cm}

\bibliographystyle{ieeetr}
\bibliography{Ref}    

\begin{thebibliography}{10}

\bibitem{1}
E.~L. Raab, M.~Prentiss, A.~Cable, S.~Chu, and D.~E. Pritchard, ``Trapping of
  neutral sodium atoms with radiation pressure,'' {\em Physical Review
  Letters}, vol.~59, no.~23, pp.~2631--2634, 1987.

\bibitem{2}
C.~Monroe, W.~Swann, H.~Robinson, and C.~Wieman, ``Very cold trapped atoms in a
  vapor cell,'' {\em Physical Review Letters}, vol.~65, no.~13, pp.~1571--1575,
  1990.

\bibitem{3}
M.~Harvey and A.~J. Murray, ``Cold atom trap with zero residual magnetic field:
  the ac magneto-optical trap,'' {\em Physical Review Letters}, vol.~101,
  p.~173201, 2008.

\bibitem{4}
T.~C. Liebisch, E.~Blanshan, E.~A. Donley, and J.~Kitching, ``Atom-number
  amplification in a magneto-optical trap via stimulated light forces,'' {\em
  Physical Review A}, vol.~85, p.~013407, 2012.

\bibitem{5}
J.~Reichel, W.~H{\"a}nsel, and T.~H{\"a}nsch, ``Atomic micromanipulation with
  magnetic surface traps,'' {\em Physical Review Letters}, vol.~83, p.~3398,
  1999.

\bibitem{6}
M.~Vangeleyn, P.~F. Griffin, E.~Riis, and A.~S. Arnold, ``Atomic
  micromanipulation with magnetic surface traps,'' {\em Optics Letters},
  vol.~35, pp.~3453--3455, 2010.

\bibitem{7}
K.~I. Lee, J.~A. Kim, H.~R. Noh, and W.~Jhe, ``Atomic micromanipulation with
  magnetic surface traps,'' {\em Optics Letters}, vol.~21, no.~15,
  pp.~1177--1179, 1996.

\bibitem{8}
R.~Stites, M.~McClimans, and S.~Bali, ``Large atom-density change at constant
  temperature by varying trap anisotropy in a dilute magneto-optical trap,''
  {\em Optics Communications}, vol.~248, pp.~173--178, 2005.

\bibitem{9}
C.~Chesman, E.~G. Lima, F.~A.~M. d.~Oliveira, S.~S. Vianna, and J.~W.~R.
  Tabosa, ``Two- and four-beam magneto-optical trapping of neutral atoms,''
  {\em Optics Letters}, vol.~19, no.~16, pp.~1237--1239, 1994.

\bibitem{10}
J.~A. Black and H.~Schmidt, ``Atomic cooling via ac stark shift,'' {\em Optics
  Letters}, vol.~39, no.~3, pp.~536--539, 2014.

\bibitem{11}
D.~J. Thomson and R.~E. Scholten, ``Narrow linewidth tunable external cavity
  diode laser using wide bandwidth filter,'' {\em Review of scientific
  instruments}, vol.~83, no.~2, p.~023107, 2012.

\bibitem{12}
S.~Svanberg, G.~Y. Yan, T.~P. Duffey, and A.~L. Schawlow, ``High-contrast
  doppler-free transmission spectroscopy,'' {\em Optics Letters}, vol.~11,
  no.~3, pp.~138--140, 1986.

\bibitem{13}
G.~L. Gattobigio, T.~Pohl, G.~Labeyrie, and R.~Kaiser, ``Scaling laws for large
  magneto-optical traps,'' {\em PHYSICA SCRIPTA}, vol.~81, p.~0253010, 2010.
\newblock The fraction of atoms in the F=3 hyperfine ground state increased by
  approximately 74\% when the atom number was raised from $1\times10^8$ to
  $4\times10^8$.

\bibitem{14}
T.~Grünzweig, A.~Hilliard, M.~McGovern, and M.~F. Andersen,
  ``Near-deterministic preparation of a single atom in an optical microtrap,''
  {\em Nature Physics}, vol.~6, pp.~951--954, 2010.

\bibitem{15}
P.~Sompet, A.~V. Carpentier, Y.~H. Fung, M.~McGovern, and M.~F. Andersen,
  ``Dynamics of two atoms undergoing light-assisted collisions in an optical
  microtrap,'' {\em Physical Review A}, vol.~88, p.~051401, 2013.

\end{thebibliography}

\clearpage

\end{document}